\documentclass[12pt,a4paper]{elsart}
\usepackage{color}
\usepackage{layout}
\usepackage{graphics}
\usepackage[dvips]{graphicx}
\usepackage{amssymb}
\usepackage{pslatex}
\def \th {\thinspace}
\journal{New Astronomy}
\begin{document}
\begin{frontmatter}
\title {The catalog of short periods stars from the ''Pi of the Sky'' data}

\author[IPJ]{A. Majczyna}, \author[OAUW]{M. Nale\.zyty},  \author[FUW]{M.
Biskup}, \author[IPJ]{G. Wrochna}, \author[IPJ]{M. Soko\l{}owski},
\author[IPJ]{K. Nawrocki}, \author[CFT]{K. Ma\l{}ek}, \author[CFT]{L. Mankiewicz},
\author[FUW]{L.W. Piotrowski}

% \ead{Agnieszka.Majczyna@fuw.edu.pl}
% \ead{nalezyty@astrouw.edu.pl}
% \ead{grzegorz.wrochna@fuw.edu.pl}

\address[IPJ]{The Andrzej So\l{}tan Institute for Nuclear Study, Ho\.za 69,
Warsaw, Poland;}
\address[OAUW]{Warsaw University Observatory, Al. Ujazdowskie 4, Warsaw, Poland;}
\address[FUW]{Faculty of Physics Warsaw University, Ho\.za 69, Warsaw, Poland;}
\address[CFT]{Center for Theoretical Physics, Polish Academy of Science, Al.
Lotnik\'ow 32/46, Warsaw, Poland;}

\begin{abstract}
Based on the data from the ''Pi of the Sky'' project we made a catalog of the
variable stars with periods from 0.1 to 10 days. We used data collected during
a period of two years (2004 and 2005) and classified 725 variable stars. Most of the
stars in our catalog are eclipsing binaries - 464 (about 64\%), while the number
of pulsating stars is 125 (about 17\%). Our classification is based on the shape of
the light curve, as in the GCVS catalog. However, some stars in our catalog were
classified as of different type than in the GCVS catalog. We have found periods for
15 stars present in the GCVS catalog with previously unknown period.
\end{abstract}
\begin{keyword} Stars: variables: catalog \end{keyword}
\end{frontmatter}
%************************************************************************************

\section{Introduction}
\label{intro}

The ''Pi of the Sky'' (Burd et al. 2005 \cite{Burd05}, and the project's web page
\cite{strona pi}) was designed and built by a team from the Andrzej So\l tan
Institute for Nuclear Study, Faculty of Physics University of Warsaw, Warsaw
University of Technology and Center for Theoretical Physics of the Polish Academy of
Science. This project applies small telescopes to observe bright objects according to
the idea proposed by prof. Bohdan Paczy\'nski (Paczy\'nski 1997 \cite{Paczynski97}). 
In 2004 thanks to Grzegorz Pojma\'nski from Warsaw University Observatory, the ''Pi
of the Sky'' project was installed in ASAS (Pojma\'nski G., 1997 \cite{Pojmanski97})
dome in Las Campanas Observatory (operated by the Carnegie Institution of
Washington).

In this work we report on observations from the first phase of the project dating
from July 2004 to December 2005. At this time the telescope was equipped with
2048$\times$2062 CCD cameras with 15$\times$15$\mu m$ pixels and 12 e$^-$ readout
noise using Carl-Zeiss telephoto lenses with  $d=f/1.4$. The robotic telescope
consists of two cameras installed on one mount. Both cameras observe the
same 33$^\circ\times$33$^\circ$ field. The telescope located at Las Campanas
Observatory could observe stars with RA\th =\th $<$\th 0\th;\th24\th $>$ hr and
% trzeba bedzie pilnowac zlamania Dec!
Dec\th=\th$<$\th-89.8$^\circ$;\th+36$^\circ$\th $>$. We do not use any filter except
for IR-cut one in order to maximize the limiting magnitude.

The main goal of the ''Pi of the Sky'' is searching for the gamma ray burst optical
afterglows so whenever it is possible, cameras track the Swift or INTEGRAL field of
view. In the remaining time  the telescope observes interesting objects from
a specially prepared list, so strong observational selection effects can be
present in the data.
% (do not use any filter).

% The main goal of the ''Pi of the Sky'' is searching for the gamma ray burst optical
% afterglows, so cameras track the Swift or INTEGRAL field of view. When it is not
% possible then the telescope observe interesting objects from the list.
% choc moze powyzszy akapit da sie wykorzystac w innym miejscu...
%************************************************************************************
\section{Data Reduction}

The data reduction was made automatically using software and scripts adapted from the
ASAS project or created specially for the ''Pi of the Sky''. Data were
divided in two streams, the first obtained directly from 10 seconds exposures and
the second from 200~s exposures obtained by co-adding of 20 images. Due to the
readout time the time resolution is 12~s and 240~s respectively. During each night a
large amount of data is produced (about 3GB/hr), so only the results are saved in our
database -- raw images are deleted after one week. The photometry is made with
different apertures -- a small one for faint stars and a large one for bright
objects. System design and observational strategy determine limiting magnitude to
about 10\th --\th 11~mag for 10\th sec exposure and $\sim$~11\th --\th 12~mag for
200~ s (Burd et al. 2005 \cite{Burd05}).

As the ''Pi of the Sky'' does not use any filter, except for the IR-cut one,
transition from the instrumental magnitude to the V filter magnitude is a source of a
systematic error. Unfortunately, value of this error is different for different
stars. The formal photometry error described as rms is equal to $\sim$~0.07~mag,
% (Ma\l{}ek 2006 \cite{Malek06}),
% 0.02 to jest rms dla fotometrii profilowej robionej przez Beate Mazur
but dispersion of the observational points in the light curve is about 0.1~mag for
stars fainter than 9~mag. A large correction is needed for different positions of
stars on the CCD due to strong vinieting and optical distorsions of lenses. To avoid
this problem one could use data taken  from one field only, but then the number of
data points drops dramatically. A visual inspection was helpful to find compromise
between the number of points and measurement quality.

The star identification in the ''Pi of the Sky'' database is based on comparison
with stars from the Tycho-2 catalog (ESA 1997 \cite{ESA97}, H\o g 1997 \cite{Hog97}).
Matching is based mainly on coordinates. Only a crude check of magnitude is performed
because of the filter correction problem mentioned above. The identification
procedure assumes that identification is positive if an investigated star is closer
than 2~acrmin to a star in the Tycho-2 catalog. An estimated error of astrometry is
about 0.5~acrmin (Biskup 2007 \cite{Biskup07}).

%************************************************************************************
%\section{Poszukiwanie zmienno¶ci}
%************************************************************************************
\section{Searching for variability}
We checked our data to find all variable stars and determine their variability
periods. All measurements for 925\thinspace 201 stars were analyzed. We used the AoV
algorithm (Schwar\-zen\-berg-Czer\-ny 1989 \cite{Schwarzenberg-Czerny89}) to
determine periods, and rejected stars with the statistic $\Theta$ larger than 50.0.
We checked periods in the range from 0.1 to 10~days for stars with a number of
observational points larger than 200. Next, some stars were rejected during the
visual inspection, so only 725 stars were classified as variable stars. The details
of this procedure are described in Biskup (2007) \cite{Biskup07}. In the next step we
determined variability type of each star.

Stars were divided in two groups (Biskup 2007 \cite{Biskup07}) - stars existing in
the ASAS catalog (Pojma\'nski 1997 \cite{Pojmanski97}) and these existing only in
 the GCVS catalog ({Kholopov et al. 1985 \cite{Kholopov85}, Kholopov et al. 1992
\cite{Kholopov92}). The automatic procedure of a period determination returned half
period instead of the full one for a significant number of stars (about 26\%). For
this reason each period was checked visually and corrected if necessary.

\begin{table}[!h]
\begin{center}
\caption{Stars and periods which were determined for the first time from our data.}
\vskip0.3cm
\label{Pgcvs}
\begin{scriptsize}
\begin{tabular}{|c|c|c|c|c|c|c|c|}
\hline
ID            & RA       & Dec       & P [d]     & Pi class& $P_{GC}$ & GCVS class&
other name  \\
\hline
000360+1208.8 & 00:03:60 &  12:08:46 & 0.1701    & DSCT     & no     & DSCTC      &
NN Peg    \\
001231+1433.8 & 00:12:31 &  14:33:51 & 1.8178:   & var      & no     & RS         &
LN Peg    \\
023602+0625.8 & 02:36:02 &  06:25:49 & 0.2079    & DSCT     & no     & DSCT       &
DX Cet    \\
052760+1254.8 & 05:27:60 &  12:54:45 & 0.3793:   & var      & no     & EB         &
V1371 Ori \\
104508+1620.1 & 10:45:08 &  16:20:07 & 0. 2043   & EW       & no     & EW         &
EX Leo    \\
114157-2423.2 & 11:41:57 & -24:23:10 & 0.1363    & DSCT     & no     & DSCTC      &
VY Crt    \\
120920-2759.3 & 12:09:20 & -27:59:18 & 0.2923    & EB       & no     & EB         &
QY Hya    \\
124420-0840.3 & 12:44:20 & -08:40:16 & 0.1167    & EA       & no     & EA/D       &
HW Vir    \\
132654-0555.7 & 13:26:54 & -05:55:40 & 0.4923    & EB:      & no     & EB         &
LU Vir    \\
141420-1521.2 & 14:14:20 & -15:21:11 & 0.2984    & DSCT/BY  & no     & BY         &
MV Vir    \\
141742-2149.6 & 14:17:42 & -21:49:37 & 0.1539    & DSCT:    & no     & DSCTC      &
MX Vir    \\
143205-2742.7 & 14:32:05 & -27:42:40 & 0.9123    & EB       & no     & EB         &
V0356 Hya \\
150401-2803.7 & 15:04:01 & -28:03:43 & 0.1466    & DSCT     & no     & DSCT       &
HY Lib    \\
173737-4048.8 & 17:37:37 & -40:48:48 & 3.3864    & DCEP     & no     & DCEPS      &
V0950 Sco \\
234535+2528.5 & 23:45:35 &  25:28:31 & 0.5790    & EW       & no     & EW         &
V0357 Peg \\
\hline
\end{tabular}
\end{scriptsize}
\end{center}
\end{table}

In Table~\ref{Pgcvs} we present a list of 15 stars with  previously unknown periods,
which were determined from our data. These stars are present in the GCVS catalog, but
without period determinations.

The classification is based on the shape of the light curve analogically to the
procedure in the GCVS catalog and in some cases based on the additional information
about a spectral type. Our strategy is different from that in the ASAS catalog, where
the classification is based on a decomposition of the light curve (Pojma\'nski 2002
\cite{Pojmanski02}).
% GW ma watpliwosci czy to co robi p. Pojmanski to dekompozycja krzywej blasku czy
% to cos moze nazywa sie inaczej
Variability types were determined simply by a visual inspection of light curves.

Symbols denoting variability types used in our classification are summarized in
Table~\ref{typy}. 
In addition, we have introduced symbols describing a situation when the
classification is ambiguous: E - eclipsing binary, var - variable star, and '':''
uncertain.
\begin{table}[!h]
\caption{Different types of variability described in our catalog of variable
stars.\vskip0.3cm}
\label{typy}
\begin{small}
\begin{center}
\begin{tabular}{|c|c|c|c|}
\hline
Symbol & Name of prototype & Period (days) & Amplitude (in V filter)\\
\hline \hline
\multicolumn{4}{|c|}{Eclipsing binaries} \\ \hline 
EA     & Algol           & 0.2 -- 1000  & $<$ several \\
EB     & $\beta$ Lyrae   & $>$ 1.0      & $<$ 2       \\
EW     & W~Ursa Maioris  & $<$ 1.0      & $<$ 0.8     \\ \hline 
\multicolumn{4}{|c|}{Pulsating stars} \\ \hline
RRAB   & RR Lyrae        & 0.3 -- 1.2   & 0.5 -- 2.0  \\
RRC    & RR Lyrae        & 0.2 -- 0.5   & $<$ 0.8     \\
DCEPS  & $\delta$ Cephei & $<$ 7.0      & $<$ 0.5     \\
BCEP   & $\beta$ Cephei  & 0.1 -- 0.6   & 0.01 -- 0.3 \\
DSCT   & $\delta$ Scuti  & 0.01 -- 0.2  & 0.003 -- 0.9\\
CW     & W Virginis      & 0.8 -- 35.0  & 0.3 -- 1.2  \\ \hline
\multicolumn{4}{|c|}{Other} \\ \hline
ACV    & Alpha2 CVn      & 0.5 -- 160.0 & 0.01 -- 0.1 \\
INT/IT & ---             & 1.0 -- 10.0  & $<$ 1.0     \\
\hline
\end{tabular}
\end{center}
\end{small}
\end{table}

\begin{figure}[!h]
\begin{center}
\includegraphics[scale=0.45]{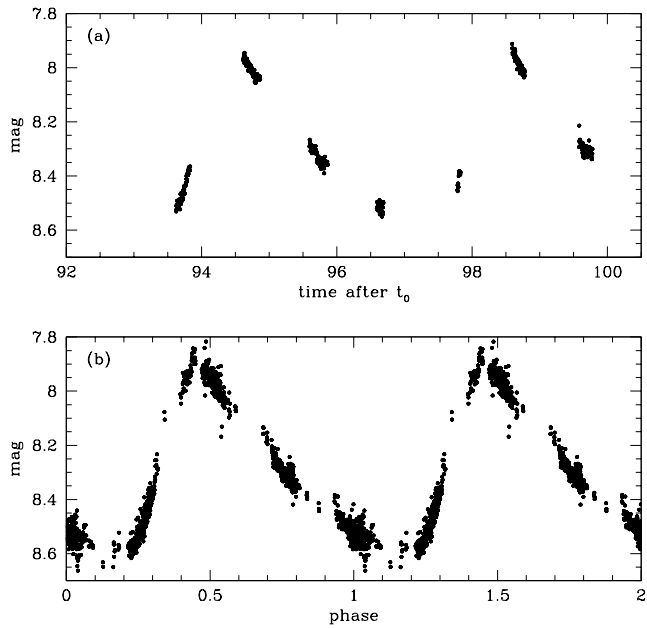}
\end{center}
\caption{{\footnotesize (a) the light curve of the ST Tau star of DCEP type, with
the period P\th=\th4.0259\th d;  (b) phased light curve }}
\label{DCEP}
\end{figure}

In case of the classification of eclipsing binaries  $\beta$~Lyrae type (EB) and
W~Ursae~Majoris (EW) we used two criteria. First, we assumed that for EW stars the
period should be shorter than 1.2\th d. Second, the secondary minimum of the light
curve should be deeper than 1/3 depth of the primary minimum. If at least one of
these conditions had not been fulfilled, such star was denoted as EB/EW.
% For pulsating star the main source of uncertainty is quality of the data. <- to
% moze nie byc prawda

\begin{figure}[!h]
\label{EW}
\begin{center}
\includegraphics[scale=0.40]{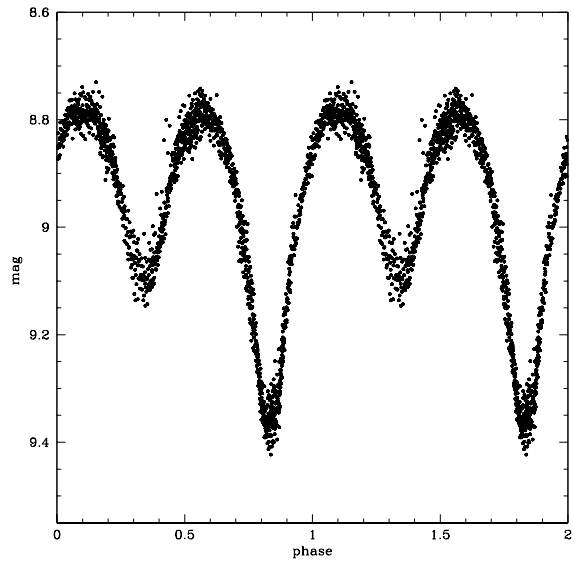}
\end{center}
\caption{{\footnotesize Phased light curve of the RV Crv star of EW~type with the
period P\th=\th0.7473\th d}}
\end{figure}

\section{Conclusions}
\subsection{Eclipsing binaries}
In Table~\ref{ile} we summarized the number of stars with a different type of
variability.
The most numerous group of stars is of the EW~type. This is in
disagreement with the GCVS catalog, where the number of EW stars is about 4 times
smaller than the number of EA stars. On the other hand, the number of EB~type stars
is almost as large as the number of EW~type stars much like in the GCVS catalog. Such
effects are artificial, resulting from assumptions made during searching for the
variability. We investigated only variable stars with periods in the range from 0.1~d
to 10 days, so for that reason there is an excess of stars with short periods in
comparison with the~GCVS catalog.

\begin{table}[!h]
\caption{Number of stars of each type from the project ''Pi of the Sky'' } \vskip
0.3cm
\label{ile}
\begin{scriptsize}
\begin{center}
\begin{tabular}{|c|c||c|c|}
\hline
Type & Number of stars & Type & Number of stars\\ \hline
EA          & 43  & RR	      &   2\\
EA:   	    & 4   & RRAB      &  36 \\
EB          & 83  & RRC	      &  11 \\
EB:         & 10  & RRC/DSCT  &  3 \\
EW          & 163 & RR/DCEP   &  1 \\
EW:         & 50  & DCEP      &  18 \\
EA/EB       &  40 & DCEPS     &  4 \\
EB/EW       &  35 & BCEP      &   1 \\
E           &  10 & DSCT      &  48 \\
E:          &  26 & BCEP/DSCT &   1 \\
EW/RR       &   1 & DSCT/BY   &   1 \\
EW/RRC      &   5 & CW	      &  5 \\
EW/DSCT     &  12 &  ACV      & 3  \\
INT/IT      &  1  & CW/DCEP   &  1 \\
EW/DSCT/RRC &  1  & var	      &  73 \\
var:        & 33  &           &\\ \hline
\end{tabular}
\end{center}
\end{scriptsize}
\end{table}

Such discrepancy disappears when we comapare our results with the ASAS catalog. Note
that Pojma\'nski (1997 \cite{Pojmanski97}) used different classification criteria
than in our work, so there is no simple relation between e.g. EW and EC stars, and a
comparison of results from these two catalogs is difficult. However, it is possible
to make some general conclusions. Table~2 of Pojma\'nski (2002) \cite{Pojmanski02}
shows that most of binary stars are contact ones - EC type, where both stars fill
their Roche lobe. In our catalog the situation is similar, the most numerous are
EW~type (W~UMa~type) binary stars. Less numerous in our catalog are EB and EA types
stars and a similar trend appears in the ASAS catalog for ESD and ED stars as well.
One should keep in mind differences in classification criteria in the ASAS
and Pi catalogs, and strong selection effects in the ''Pi of the Sky'' data.
\begin{figure}[!h]
\label{RRAB}
\begin{center}
\includegraphics[scale=0.35]{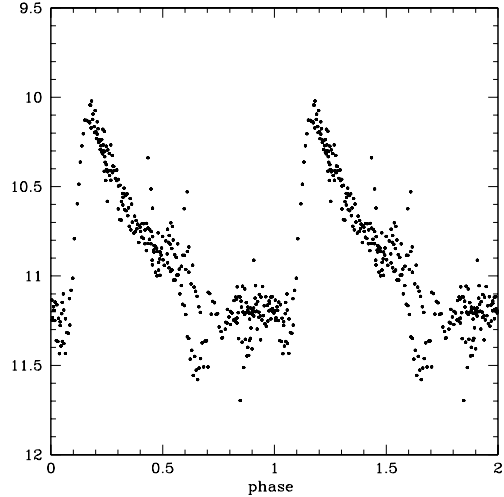}
\end{center}
\caption{{\footnotesize Phased light curve of the RR Leo, RRAB type star with the
period P=0.4520d }}
\end{figure}

\subsection{Pulsating stars}
Discrepancy between the number of different types of pulsating stars in the ''Pi of
the Sky'' and GCVS catalogs can not be explained by the simple fact that we
investigated only stars with periods shorter than 10 days. All pulsating stars
present in our catalog have periods much shorter than 10 days. For example, the ratio
of the number of RRAB stars to the number of all stars in our catalog is 0.05, while
in the GCVS catalog it is equal to 0.11. Comparing the ratio of numbers of RRAB and
DSCT type stars, we obtained the following values: ASAS 1.7 (based on Pojma\'nski
2002 \cite{Pojmanski02}), ''Pi of the Sky'' about 0.75, and GCVS 39.6. Similar
situation occurs for other types of pulsating stars as well. The main difference
between the ''Pi of the Sky'' or ASAS and GCVS catalogs is the magnitude range. In
the GCVS catalog there are informations about stars with -1.4\th --\th 20 mag, while
the magnitude range for ''Pi of the Sky'' is 5\th --\th 12~mag and for ASAS 8.5\th
--\th 15~mag. We interpreted differences in the number of each type of stars in these
3 catalogs as an effect of a different magnitude range in theses projects and effect
of a specific observational strategy in the ''Pi of the Sky'' project.
%''Pi'' do
% not made a systematic scan of the sky, but track field of view of the gamma ray
% satelites, so there is strong observational effect.

%************************************************************************************
\section{Summary and future perspective}
We presented the catalog of 725 variable stars from the first phase of the ''Pi of
the Sky'' project. The catalog contains stars with periods ranging beetwen 0.1 and
10~days. The types of variability were determined through a visual inspection.\\
Most of the variable stars in our catalog are eclipsing binaries, W UMa type. Among
the pulsating stars most of them are $\delta$ Scuti type stars (48), but
RRAB stars are only slightly less numerous (36). Both type of stars have rather large
variability amplitude, so are easy to detect.
% For 15 stars known from GCVS catalog (none in ASAS) we determined periods.
We determine accurate periods for 15 stars from the GCVS catalog with periods unknown
so far.

The catalog and whole data base of measurements are available at the ''Pi of the
Sky''
website \verb+http://grb.fuw.edu.pl/pi/index.html#star_catalog+

Since the primary goal of the ''Pi of the Sky'' is searching for the GRB prompt
optical emission, the algorithm has been optimized for the flash recognition.
We work on improving the photometry and we plan to develop an automatic procedure for
a classification of variable stars. This will be necessary for the final version of
the ''Pi of the Sky'' project where the amount of data will increase by almost two
orders of magnitude.

% \acknowledgments
\vskip0.5cm
{\bf Acknowledgments}. We are very grateful to Grzegorz Pojma\'nski for access to
the ASAS dome and many practical advices.\\
We would like to thank the staff of the Las Campanas Observatory for their help
during the installation and maintenance of our telescope.\\
This work was financed by the Polish Ministry of Science in 2005-2007 as a research
project. This work is also supported  by the Polish Ministry of Science and Higher
Educattion grant No. N\th N203\th 4061\th 33 (AM, MN).

%************************************************************************************

% \include{kat_fina1.3}
% na wyrazna sugestie/zadanie tabela z katalogiem zostaje usunieta z tekstu

\begin {thebibliography}{99}
\bibitem {Biskup07}  Biskup, M. 2007, ''Variable stars search in the Pi of the Sky
experiment'' Master's thesis, Faculty of Physics Warsaw University
\bibitem {Burd05} Burd, A. et al. 2005, New Astronomy 10, 409
\bibitem {ESA97} ESA 1997, The Hipparcos and Tycho Catalogues, ESA SP-1200
\bibitem {Hog97} H\o g E., 1997 Proceedings of ESA Symposium 'Hipparcos - Venice'97',
ESA-SP 402
\bibitem {strona pi} http://grb.fuw.edu.pl
\bibitem {Kholopov85} Kholopov, P.N. et al. 1985-1988, General Catalogue of Variable
Stars 4rd ed. vols. I-III, Nauka, Moscow
\bibitem {Kholopov92}  Kholopov, P.N. et al.1992, General Catalogue of Variable Stars
4rd ed. vol. IV, Bull. Inf. CDS 40, 15
\bibitem {Paczynski97} Paczy\'nski B., 1997 The Future of Massive Variability
Searches, in Variable Stars and the Astrophysical Returns of the Microlensing
Surveys. Edited by Roger Ferlet, Jean-Pierre Maillard and Brigitte Raban. Cedex,
France: Editions Frontieres, 1997, p.357. %(astro-ph: 9609073)
\bibitem {Pojmanski97}  Pojma\'nski; G. 1997, Acta Astronomica 47, 467
% jesli zastosuje \bibitem[]{} to bedzie [1] Majczyna et al.
\bibitem {Pojmanski02} Pojma\'nski G. 2002, Acta Astronomica 52, 397
\bibitem{Schwarzenberg-Czerny89}  Schwarzenberg-Czerny, A. 1989, MNRAS 241, 153
\end{thebibliography}
\end{document}